\documentclass[twocolumn,amssymb,amsmath,aps,showpacs,nofootinbib,nobibnotes,prl]{revtex4-1}

\usepackage{graphicx}
\usepackage{xcolor}
\usepackage[colorlinks=true,linkcolor=blue,urlcolor=blue,citecolor=blue,bookmarks=true,bookmarksopenlevel=2]{hyperref}
\usepackage{amsmath}

\usepackage[utf8]{inputenc}
\usepackage[T1]{fontenc}
\usepackage{microtype}

\newcommand{\alert}[1]%
{%
\textcolor{red}{#1}
}%

\begin{document}

\title{Relativistic calculated K-shell level widths and fluorescence yields\\for atoms with 20$\le$Z$\le$30}
\author{Karol Kozio{\l}$^{1,2}$} 
\email{mail@karol-koziol.net}
\affiliation{$^1$Faculty of Chemistry, Nicolaus Copernicus University, 87-100 Toru\'n, Poland}
\affiliation{$^2$Natural and Exact Science Faculty, Northeastern University of Argentina, W3404AAS Corrientes, Argentina}

\begin{abstract}
The $K$-shell level radiative, non-radiative, and total widths, $K$-shell fluorescence yields and $K$-shell hole state lifetimes for atoms with 20$\le$Z$\le$30 have been calculated in fully-relativistic way using the extensive multiconfiguration Dirac-Fock calculations with the inclusion of the Breit interaction and QED corrections and also using multiconfiguration Dirac-Hartree-Slater calculations. 
\end{abstract}


\maketitle

\section{Introduction}

$K$-shell atomic level width, $\Gamma_K$, and fluorescence yields, $\omega_K$, have been an object of the intense research for several decades, both experimentally and theoretically, and a large number of articles have been published on these subjects. 
%
In 1969-70, McGuire \cite{mcguire69,mcguire70} was the first who published $\omega_K$ values for atoms with 4$\le$Z$\le$54. Then, in 1971, Walters and Balla \cite{walters} and Kostroun et al. \cite{kostroun} published $\omega_K$ values for atoms with 5$\le$Z$\le$54 and 10$\le$Z$\le$70, respectively. 
These calculations were based on the Hartree-Fock-Slater method. 
In 1972, Bambynek et al. \cite{bambynek72} published a review article with a selection of the "most reliable" experimental and theoretical $\omega_K$ values for selected atoms with 13$\le$Z$\le$92. 
In 1979, Krause and Oliver published compilation of $\Gamma_K$ \cite{krause} and $\omega_K$ \cite{krause-omegi} values for atoms with 5$\le$Z$\le$110, basing on both experimental and theoretical data. 
In 1980, Chen et al. \cite{chen80} published $\Gamma_K$ and $\omega_K$ values for selected atoms with 18$\le$Z$\le$96, calculated by using relativistic Dirac-Hartree-Slater method. 
Afterwards, in 1994, Hubbell et al. \cite{hubbell} published a compilation of experimental and semi-empirical $\omega_K$ values for atoms with 3$\le$Z$\le$110. 
In 2001, Campbell and Papp presented \cite{papp} a compilation of recommended $\Gamma_K$ values for atoms with 10$\le$Z$\le$92 basing on experimental data and earlier compilation of EADL \cite{eadl} from 1991. 
Although these data are already several decades old and in the last years the more advanced relativistic methods have been developed, they are still widely used for comparison with experimental results.

The knowledge of natural $K$-shell level widths are a base for investigation of width of various $K$-shell X-ray lines; 
the comparison between theoretical and experimental X-ray linewidths leads to analysis of ionization mechanisms in the inner shells of atom (see eg. \cite{lifetimes-prl}).  
A good knowledge of $K$-shell fluorescence yields is important for the analysis of various measurements in the fields of nuclear, atomic, molecular, and applied physics (see eg. \cite{bambynek72}). 
On the other hand, in the last years, there have been many experimental high precision measurements of the $K$-shell fluorescence yields for atoms with 20$\le$Z$\le$30 \cite{yashoda,sahin,gudennavar,durak,simsek,sogut,han}, but the full agreement between experimental and theoretical or semi-empirical data was not obtained. 

In this work the theoretical predictions for radiative, non-radiative, and total natural $K$-shell level widths, as well as $K$-shell fluorescence yields and $K$-shell hole state lifetimes, for atoms with 20$\le$Z$\le$30 have been evaluated basing on the multiconfiguration Dirac-Fock (MCDF) method and the multiconfiguration Dirac-Hartree-Slater (DHS) method. 
It is important to note that in case of open-shell atoms (when de-excitation concerns many states) every state has a specific width, thus, radiative and total widths presented in this work are the average values evaluated as an arithmetic mean of all widths for individual states. 

\section{Theoretical background}

\subsection{Relativistic calculations}

The methodology of MCDF calculations performed in the present studies is similar to the published earlier, in many papers (see, e.g., \cite{gr1,gr2,gr3,gr4,gr5,gr6,grant1,grant2,polasik1989a,polasik1995}).
The effective Hamiltonian for an N-electron system is expressed by
\begin{equation}
H = \sum_{i=1}^{N} h_{D}(i) + \sum_{j>i=1}^{N} C_{ij},
\end{equation}
where $h_{D}(i)$ is the Dirac operator  for $i$-th electron and the term $C_{ij}$ accounts for electron-electron interactions. 
The latter is a sum of the Coulomb interaction operator and the transverse Breit operator.
An atomic state function (ASF) with the total angular momentum $J$ and parity $p$ is assumed in the form
\begin{equation}
\Psi_{s} (J^{p} ) = \sum_{m} c_{m} (s) \Phi ( \gamma_{m} J^{p} ),
\end{equation}
where $\Phi ( \gamma_{m} J^{p} )$ is a configuration state function (CSF), $c_{m} (s)$ is a configuration mixing coefficient for state $s$, $\gamma_{m}$ represents all information  required  to define a certain CSF uniquely.
Apart from the transverse Breit interaction two types of quantum electrodynamics (QED) corrections have been included (self-energy and vacuum polarization). 

In general, the multiconfiguration DHS method is similar to the MCDF method, but simplified expression for electronic exchange integrals is used \cite{fac2}. 

\begin{table*}[!htb]
\caption{\label{tab:states}Numbers of $K^{-1}$, $L_{2,3}^{-1}$, $M_{2,3}^{-1}$, $L^{-2}$, $L^{-1}M^{-1}$, $L^{-1}N^{-1}$, $M^{-2}$, $M^{-1}N^{-1}$ for considered valence electronic configurations. }
\medskip
\begin{tabular*}{\linewidth}{@{\extracolsep{\fill}}ccc ccc ccccc}
\hline\\[-1.5ex]
Z&Atom&Valence&\multicolumn{8}{c}{Number of states for spectator hole(s)}\\
\cline{4-11}\\[-2ex]
~&~&configuration& $K^{-1}$ & $L_{2,3}^{-1}$ & $M_{2,3}^{-1}$ & $L^{-2}$ & $L^{-1}M^{-1}$ & $L^{-1}N^{-1}$ & $M^{-2}$ & $M^{-1}N^{-1}$ \\[0.5ex]\hline\\[-1.5ex]
20 & Ca & $4s^2$ & 1 & 2 & 2 & 10 & 20 & 6 & 10 & 6 \\
21 & Sc & $3d^{1}4s^2$ & 4 & 12 & 12 & 53 & 122 & 31 & 56 & 32 \\
22 & Ti & $3d^{2}4s^2$ & 16 & 45 & 45 & 210 & 484 & 122 & 227 & 126 \\
23 & V  & $3d^{3}4s^2$ & 38 & 110 & 110 & 493 & 1174 & 287 & 556 & 303 \\
24 & Cr & $3d^{5}4s^1$ & 144 & 417 & 417 & 1934 & 4858 & 288 & 2458 & 322 \\
25 & Mn & $3d^{5}4s^2$ & 74 & 214 & 214 & 967 & 2429 & 561 & 1229 & 624 \\
26 & Fe & $3d^{6}4s^2$ & 63 & 180 & 180 & 838 & 2172 & 486 & 1160 & 560 \\
27 & Co & $3d^{7}4s^2$ & 38 & 110 & 110 & 493 & 1356 & 287 & 773 & 350 \\
28 & Ni & $3d^{8}4s^2$ & 16 & 45 & 45 & 210 & 616 & 122 & 392 & 160 \\
29 & Cu & $3d^{10}4s^1$ & 2 & 4 & 4 & 16 & 66 & 3 & 63 & 5 \\
30 & Zn & $3d^{10}4s^2$ & 1 & 2 & 2 & 10 & 36 & 6 & 35 & 10 \\[0.5ex]
\hline
\end{tabular*}
\end{table*}

\begin{table*}[!htb]
\caption{\label{tab:krad1}Radiative level widths of $K^{-1}$ hole state atomic level for atoms with 20$\le$Z$\le$30 (Coulomb gauge).}
\medskip
\begin{tabular*}{\linewidth}{@{\extracolsep{\fill}}cc cc cc}
\hline\\[-1.5ex]
Z&Atom&\multicolumn{2}{c}{$\bar X_i$ [s$^{-1}$]}&$\sum_i \bar X_i$  &$\Gamma^{Rad}$\\
\cline{3-4}\\[-2ex]
~&~&$K\alpha_{1,2}$ [$\times10^{14}$] & $K\beta_{1,3}$ [$\times10^{13}$] & [$\times10^{14}$~s$^{-1}$] &[eV]\\[0.5ex]\hline\\[-1.5ex]
20 & Ca & 1.737 & 2.412 & 1.978 & 0.130 \\
21 & Sc & 2.167 & 2.827 & 2.449 & 0.161 \\
22 & Ti & 2.670 & 3.531 & 3.023 & 0.199 \\
23 & V & 3.257 & 4.351 & 3.692 & 0.243 \\
24 & Cr & 3.935 & 5.142 & 4.450 & 0.293 \\
25 & Mn & 4.712 & 6.391 & 5.351 & 0.352 \\
26 & Fe & 5.599 & 7.635 & 6.362 & 0.419 \\
27 & Co & 6.605 & 9.047 & 7.509 & 0.494 \\
28 & Ni & 7.739 & 10.643 & 8.804 & 0.579 \\
29 & Cu & 9.015 & 11.007 & 10.116 & 0.666 \\
30 & Zn & 10.424 & 12.950 & 11.719 & 0.771 \\[0.5ex]
\hline
\end{tabular*}
\end{table*}

\begin{table*}[!htb]
\caption{\label{tab:krad2}Radiative level widths of $K^{-1}$ hole state atomic level for atoms with 20$\le$Z$\le$30 (Babushkin gauge).}
\medskip
\begin{tabular*}{\linewidth}{@{\extracolsep{\fill}}cc cc cc}
\hline\\[-1.5ex]
Z&Atom&\multicolumn{2}{c}{$\bar X_i$ [s$^{-1}$]}&$\sum_i \bar X_i$  &$\Gamma^{Rad}$\\
\cline{3-4}\\[-2ex]
~&~&$K\alpha_{1,2}$ [$\times10^{14}$] & $K\beta_{1,3}$ [$\times10^{13}$] & [$\times10^{14}$~s$^{-1}$] &[eV]\\[0.5ex]\hline\\[-1.5ex]
20 & Ca & 1.579 & 2.758 & 1.855 & 0.122 \\
21 & Sc & 1.985 & 3.202 & 2.305 & 0.152 \\
22 & Ti & 2.459 & 3.974 & 2.856 & 0.188 \\
23 & V & 3.012 & 4.868 & 3.499 & 0.230 \\
24 & Cr & 3.654 & 5.726 & 4.227 & 0.278 \\
25 & Mn & 4.391 & 7.074 & 5.098 & 0.336 \\
26 & Fe & 5.234 & 8.411 & 6.075 & 0.400 \\
27 & Co & 6.192 & 9.921 & 7.184 & 0.473 \\
28 & Ni & 7.274 & 11.622 & 8.437 & 0.555 \\
29 & Cu & 8.491 & 12.021 & 9.694 & 0.638 \\
30 & Zn & 9.814 & 12.859 & 11.100 & 0.731 \\[0.5ex]
\hline
\end{tabular*}
\end{table*}

\subsection{Lifetime of excited states, width of corresponding atomic levels, and fluorescence yields\label{sec:teor-lif-wid}}

Each excited state can be attributed by the mean lifetime~$\tau$. The mean lifetime can be defined as time after which the number of excited states of atoms decreases $e$ times. The mean lifetime is determined by a total transition rate of de-excitation (radiative and non-radiative) processes $W_i$: 
\begin{equation}
\tau = \left(\sum_i W_i\right)^{-1} = \left(\sum_i X_i + \sum_j A_j + \sum_k C_k\right)^{-1}\;,
\label{eq:tau}
\end{equation} 
where $X_i$ is a transition rate of radiative process, $A_j$ is a transition rate of non-radiative Auger process, and $C_k$ is a transition rate of non-radiative Coster-Kronig or super-Coster-Kronig process. De-excitation processes happen in all possible ways leading to lower energetic states allowed by selection rules.

Due to the energy-time uncertainty principle ($\Delta E \Delta t = \hbar$), the lifetime of excited state $\tau$ is connected with width of corresponding atomic level $\Gamma$ (note that for one atomic level there may be more than one corresponding excited states) 
by the relationship
\begin{equation}
\Gamma = \frac{\hbar}{\tau} = \hbar W = \hbar \sum_i W_i\;. 
\label{eq:gam}
\end{equation}
Natural width of atomic level can be obtained in form of a sum of radiative width $\Gamma_X$, Auger width $\Gamma_A$, and Coster-Kronig $\Gamma_C$ width; or, in other way, as a sum of radiative and non-radiative widths:
\begin{equation}
\Gamma = \Gamma_X + \Gamma_A +\Gamma_C = \Gamma^{Rad} + \Gamma^{Nrad}\;,
\label{eq:gamma}
\end{equation}
where $\Gamma^{Rad} = \Gamma_X$ and $\Gamma^{Nrad} = \Gamma_A +\Gamma_C$. 
The relevant yields are linked to the terms presented above, i.e.
\begin{subequations}
\begin{eqnarray}
&\omega = \frac{\Gamma_X}{\Gamma} = \frac{\sum_i X_i}{\sum_i X_i + \sum_j A_j + \sum_k C_k}\;,&\\
&a = \frac{\Gamma_A}{\Gamma} = \frac{\sum_j A_j}{\sum_i X_i + \sum_j A_j + \sum_k C_k}\;,&\\
&f = \frac{\Gamma_C}{\Gamma} = \frac{\sum_k C_k}{\sum_i X_i + \sum_j A_j + \sum_k C_k}\;,&
\label{eq:omega}
\end{eqnarray}
\end{subequations}
where $\omega$ is fluorescence yield, $a$ is Auger yield, and $f$ is Coster-Kronig yield.

For open-shell atomic systems for each atomic hole level $(nlj)^{-\alpha}$ (where $\alpha$= 1, 2, \ldots, $2j+1$) a great number of hole states $(nlj)_J{}^{-\alpha}$ are linked. 
Therefore, for $i$-th hole state all transition rates $W_{ij}$ corresponded to $j$-th de-excitation process should be considered. 
Radiative part of natural width $\Gamma^{Rad}_i$ of $i$-th hole state can be determined by using transition rate of radiative processes $X_{ij}$ according to the formula 
\begin{equation}
\Gamma^{Rad}_i = \hbar \sum_j X_{ij} \;.
\label{eq:grad}
\end{equation}
Radiative part of the natural width of $(nlj)^{-\alpha}$ hole atomic level is taken as the arithmetic mean of the radiative parts to the natural width of the each hole state $(nlj)_J{}^{-\alpha}$, i.e. according to the following formula 
\begin{equation}
\Gamma^{Rad} = \frac{\sum_i \Gamma^{Rad}_i}{n} \;,
\label{eq:ave}
\end{equation}
where $\Gamma^{Rad}$ is a radiative part of natural level width, and $n$ is a number of hole state corresponding to a given hole level. 

Similarly, the non-radiative part of the natural width can be determined by calculating the transition rates for non-radiative Auger $A_{ij}$ and Coster-Kronig $C_{ik}$ processes according to the formula 
\begin{equation}
\Gamma^{Nrad} = \frac{\hbar \left(\sum_{i,j} A_{ij} + \sum_{i,k} C_{ik}\right)}{n} \;,
\label{eq:gnrad}
\end{equation}
where all the designations are analogous to those above. 
Then the total natural width of the atomic hole levels can be determined according to the Eq.~\ref{eq:gamma}.

Equations \ref{eq:grad}, \ref{eq:ave}, and \ref{eq:gnrad} can be rewritten to the form of:
\begin{equation}
\Gamma^{Nrad} = \hbar \sum_{j} \bar X_{j} \;,
\label{eq:grada}
\end{equation}
\begin{equation}
\Gamma^{Nrad} = \hbar \left(\sum_{j} \bar A_{j} + \sum_{k} \bar C_{k}\right) \;,
\label{eq:gnrada}
\end{equation}
where $\bar X_j = \frac{1}{n} \sum_{i} X_{ij}$ is a mean value of transition rates per one $(nlj)_J{}^{-\alpha}$ hole state for $j$-th de-excitation channel, and for $\bar A$ and $\bar C$ similarly. 

\section{Results and discussion}

Considered radiative de-excitation processes for $K^{-1}$ hole states are: 
$K\alpha_{1,2}$ ($K$-$L_{2,3}$) and $K\beta_{1,3}$ ($K$-$M_{2,3}$). 
These are electric (E1) dipole transitions. Transitions of higher order (E2, M1, M2, etc.) were not included here due to the small contribution to the level widths. 
Considered radiative de-excitation processes are: $K$-$LL$, $K$-$LM$, $K$-$LN$, $K$-$MM$, and $K$-$MN$. 
It is worth noticing that in the case of $K^{-1}$ hole states there is no Coster-Kronig de-excitation process. 
For open-shell atoms, numerous states are engaged for these de-excitation channels (see Table~\ref{tab:states}).

The calculations of radiative transition rates were carried out by means of \textsc{Grasp2k} code \cite{grasp2k}. The radiative transition rates were calculated in both Coulomb \cite{grant1} and Babushkin \cite{cech-bab} gauges. 
The calculations of non-radiative transition rates were carried out by means of both \textsc{Fac} code \cite{fac1,fac2} and a piece of code from \textsc{Ratip} suite \cite{ratip,ratip2012}, which uses \text{Grasp92} \cite{grasp92} calculated wavefunctions. 
All the versions of \textsc{Grasp} code are fully MCDF codes. The \textsc{Fac} code is based on modified DHS method, but multiconfigurational approach and Breit interaction are also included.

\begin{table*}[!htb]
\caption{\label{tab:knrad1}Non-radiative level widths of $K^{-1}$ hole state atomic level for atoms with 20$\le$Z$\le$30 (\textsc{Fac} calculated).}
\medskip
\begin{tabular*}{\linewidth}{@{\extracolsep{\fill}}cc ccccc cc}
\hline\\[-1.5ex]
Z&Atom&\multicolumn{5}{c}{$\bar A_i$ [s$^{-1}$]}&$\sum_i \bar A_i$  &$\Gamma^{Nrad}$\\
\cline{3-7}\\[-2ex]
~&~& K-LL & K-LM & K-LN & K-MM & K-MN & [$\times10^{15}$~s$^{-1}$] &[eV]\\
~&~& [$\times10^{14}$] & [$\times10^{14}$] & [$\times10^{12}$] & [$\times10^{13}$] & [$\times10^{11}$] & ~ &~\\[0.5ex]\hline\\[-1.5ex]
20 & Ca & 7.744 & 1.677 & 4.388 & 0.918 & 5.318 & 0.956 & 0.629\\
21 & Sc & 8.022 & 1.774 & 4.364 & 0.990 & 5.388 & 0.994 & 0.655\\
22 & Ti & 8.282 & 1.862 & 4.319 & 1.054 & 5.411 & 1.030 & 0.678\\
23 & V & 8.525 & 1.944 & 4.264 & 1.113 & 5.404 & 1.063 & 0.700\\
24 & Cr & 8.761 & 1.983 & 1.844 & 1.123 & 2.316 & 1.088 & 0.716\\
25 & Mn & 8.965 & 2.095 & 4.137 & 1.222 & 5.340 & 1.123 & 0.739\\
26 & Fe & 9.165 & 2.166 & 4.072 & 1.274 & 5.296 & 1.150 & 0.757\\
27 & Co & 9.354 & 2.234 & 4.007 & 1.323 & 5.247 & 1.177 & 0.774\\
28 & Ni & 9.532 & 2.300 & 3.943 & 1.372 & 5.197 & 1.201 & 0.791\\
29 & Cu & 9.705 & 2.340 & 2.164 & 1.389 & 2.972 & 1.221 & 0.804\\
30 & Zn & 9.861 & 2.428 & 3.821 & 1.468 & 5.096 & 1.248 & 0.821\\[0.5ex]
\hline
\end{tabular*}
\end{table*}

\begin{table*}[!htb]
\caption{\label{tab:knrad2}Non-radiative level widths of $K^{-1}$ hole state atomic level for atoms with 20$\le$Z$\le$30 (\textsc{Ratip} calculated).}
\medskip
\begin{tabular*}{\linewidth}{@{\extracolsep{\fill}}ccc ccc cc}
\hline\\[-1.5ex]
Z&Atom&Valence&\multicolumn{3}{c}{$\bar A_i$ [s$^{-1}$]}&$\sum_i \bar A_i$  &$\Gamma^{Nrad}$\\
\cline{4-6}\\[-2ex]
~&~&configurations& K-LL & K-LM & K-MM & [$\times10^{15}$~s$^{-1}$] &[eV]\\
~&~&~& [$\times10^{14}$] & [$\times10^{14}$] & [$\times10^{13}$] & ~ &~\\[0.5ex]\hline\\[-1.5ex]
20 & Ca & $4s^2$ & 8.155 & 1.944 & 1.167 & 1.022 & 0.672 \\[0.5ex]
21 & Sc & $3d_{3/2}^1 4s^2$ & 8.356 & 2.064 & 1.278 & 1.055 & 0.694 \\[0.5ex]
22 & Ti & $3d_{3/2}^4$ & 8.657 & 2.045 & 1.209 & 1.082 & 0.712 \\[0.5ex]
23 & V & $3d_{3/2}^4 4s^1$ & 8.834 & 2.196 & 1.362 & 1.117 & 0.735 \\[0.5ex]
24 & Cr & $3d_{5/2}^6$ & 9.037 & 2.251 & 1.395 & 1.143 & 0.752 \\[0.5ex]
25 & Mn & $3d_{5/2}^6 4s^1$ & 9.204 & 2.375 & 1.521 & 1.173 & 0.772 \\[0.5ex]
26 & Fe & $3d_{5/2}^6 4s^2$ & 9.373 & 2.511 & 1.660 & 1.205 & 0.793 \\[0.5ex]
27 & Co & $3d_{3/2}^1 3d_{5/2}^6 4s^2$ & 9.503 & 2.512 & 1.734 & 1.219 & 0.802 \\[0.5ex]
28 & Ni & $3d^{10}$ & 9.693 & 2.533 & 1.642 & 1.239 & 0.816 \\[0.5ex]
29 & Cu & $3d^{10}4s^1$ & 9.830 & 2.628 & 1.738 & 1.263 & 0.831 \\[0.5ex]
30 & Zn & $3d^{10}4s^2$ & 9.969 & 2.813 & 1.930 & 1.297 & 0.854 \\[0.5ex]
\hline
\end{tabular*}
\end{table*}

\begin{table*}[!htb]
\caption{\label{tab:kwid}Total level widths of $K^{-1}$ hole state atomic level, and fluorescence yields and lifetimes of $K^{-1}$ hole states for atoms with 20$\le$Z$\le$30 for four calculation models (see text for details).}
\medskip
\begin{tabular*}{\linewidth}{@{\extracolsep{\fill}}cc cccc cccc cccc}
\hline\\[-1.5ex]
Z & Atom & \multicolumn{4}{c}{$\Gamma$ [eV]} & \multicolumn{4}{c}{$\omega$} & \multicolumn{4}{c}{$\tau$ [$\times10^{-16}$~s]}\\[0.5ex]
\cline{3-6}\cline{7-10}\cline{11-14}\\[-2ex]
 &  & C+F & B+F & C+R & B+R & C+F & B+F & C+R & B+R & C+F & B+F & C+R & B+R \\[0.5ex]\hline\\[-1.5ex]
20 & Ca & 0.760 & 0.751 & 0.803 & 0.795 & 0.171 & 0.162 & 0.162 & 0.154 & 8.666 & 8.759 & 8.201 & 8.284 \\
21 & Sc & 0.816 & 0.806 & 0.855 & 0.846 & 0.198 & 0.188 & 0.188 & 0.179 & 8.069 & 8.164 & 7.694 & 7.780 \\
22 & Ti & 0.877 & 0.866 & 0.911 & 0.900 & 0.227 & 0.217 & 0.218 & 0.209 & 7.507 & 7.602 & 7.222 & 7.311 \\
23 & V & 0.943 & 0.930 & 0.978 & 0.965 & 0.258 & 0.248 & 0.248 & 0.239 & 6.983 & 7.079 & 6.730 & 6.819 \\
24 & Cr & 1.009 & 0.994 & 1.045 & 1.030 & 0.290 & 0.280 & 0.280 & 0.270 & 6.524 & 6.620 & 6.299 & 6.388 \\
25 & Mn & 1.091 & 1.075 & 1.124 & 1.108 & 0.323 & 0.312 & 0.313 & 0.303 & 6.031 & 6.125 & 5.854 & 5.942 \\
26 & Fe & 1.176 & 1.157 & 1.212 & 1.193 & 0.356 & 0.346 & 0.346 & 0.335 & 5.597 & 5.688 & 5.431 & 5.517 \\
27 & Co & 1.269 & 1.247 & 1.297 & 1.275 & 0.390 & 0.379 & 0.381 & 0.371 & 5.188 & 5.277 & 5.077 & 5.162 \\
28 & Ni & 1.370 & 1.346 & 1.395 & 1.371 & 0.423 & 0.413 & 0.415 & 0.405 & 4.804 & 4.890 & 4.718 & 4.802 \\
29 & Cu & 1.469 & 1.442 & 1.497 & 1.469 & 0.453 & 0.443 & 0.445 & 0.434 & 4.479 & 4.566 & 4.396 & 4.479 \\
30 & Zn & 1.593 & 1.552 & 1.625 & 1.585 & 0.484 & 0.471 & 0.475 & 0.461 & 4.133 & 4.241 & 4.050 & 4.154 \\[0.5ex]
\hline
\end{tabular*}
\end{table*}

In Tables~\ref{tab:krad1} and~\ref{tab:krad2} the present calculated values of radiative part of natural $K^{-1}$ level width for atoms with 20$\le$Z$\le$30 are collected (in both Coulomb and Babushkin gauges).
In Tables~\ref{tab:knrad1} and~\ref{tab:knrad2} the present calculated  values of non-radiative part of natural $K^{-1}$ level width are collected (calculated with the use of both \textsc{Fac} and \textsc{Ratip} codes).
In order to simplify the calculations and because of the restriction in number of simultaneously optimized states within \textsc{Grasp92} code, in the case of non-radiative transition rates calculated by \textsc{Ratip} code the valence electronic configurations were simplified in order to minimize number of states. In fact, this approach can change values of non-radiative transition rates a little, but the tests show that it do not change the total value of non-radiative part of natural $K^{-1}$ level width substantially. 
The $K$-$LN$ and $K$-$MN$ non-radiative de-excitation channels are also omitted in this case, because of their minor contribution to the level width. 
For each of Tables \ref{tab:krad1}, \ref{tab:krad2}, \ref{tab:knrad1}, and \ref{tab:knrad2}, the values of present calculated radiative and non-radiative transition rates for considered de-excitation processes for $K^{-1}$ hole states are also shown in the form of mean values per one $K^{-1}$ hole state. 

The total $K^{-1}$ level widths, the fluorescence yields and the lifetimes of $K^{-1}$ states for four calculation models are collected in Table~\ref{tab:kwid}. 
The first model, marked as C+F, uses Coulomb-gauge values for radiative transitions and \textsc{Fac} calculated values for non-radiative transitions. 
The second one, marked as B+F, uses Babushkin-gauge values for radiative transitions and \textsc{Fac} calculated values for non-radiative transitions. 
The third model, marked as C+R, uses Coulomb-gauge values for radiative transitions and \textsc{Ratip} calculated values for non-radiative transitions. 
The fourth model, marked as B+R, uses Babushkin-gauge values for radiative transitions and \textsc{Ratip} calculated values for non-radiative transitions. 

One can see from Tables~\ref{tab:krad1} and~\ref{tab:krad2} and from Tables~\ref{tab:knrad1} and~\ref{tab:knrad2} that there is a~few percentage difference between radiative width values calculated by using Coulomb and Babushkin gauge; similarly, there is also a~few percentage difference between non-radiative width values calculated by using MCDF (\textsc{Ratip} code) and DHS (\textsc{Fac} code). 
As a~result, the values of total $K^{-1}$ level widths for mentioned above models differ from each other by a~few percent. 
This implicates the statement, that values of {ab initio} calculations of level widths depend not only on the advancement of computational method (MCDF vs. DHS), but also on the choice of gauge for radiative de-excitation rates. 

\begin{figure*}[!htb]
\centering
\includegraphics[width=0.85\textwidth]{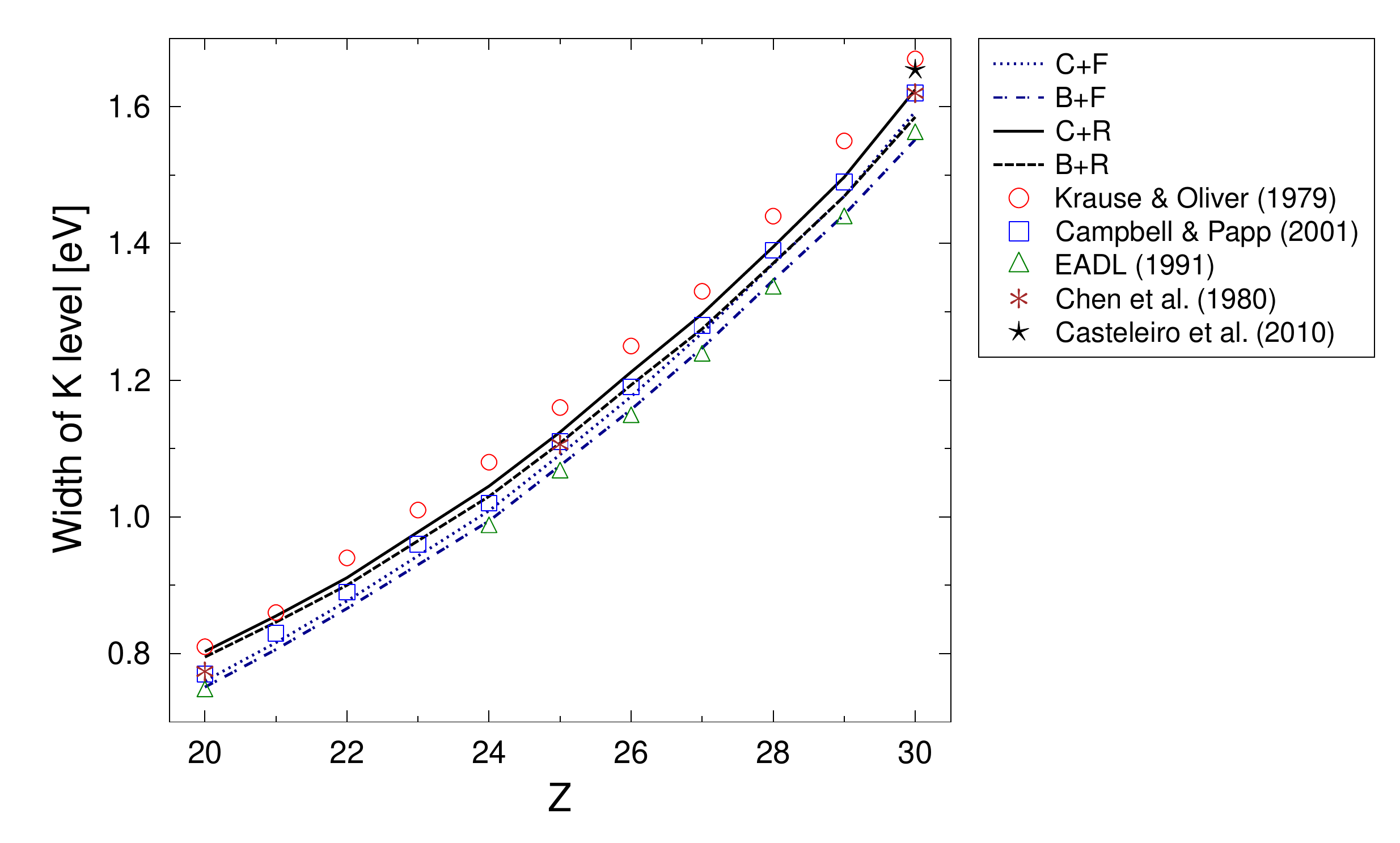}
\caption{\label{fig:kwidpor}(Color online) Level widths of $K^{-1}$ hole state atomic level for atoms with 20$\le$Z$\le$30 compared to available theoretical and semi-empirical literature data: Krause and Oliver \cite{krause}, Campbell and Papp \cite{papp}, EADL \cite{eadl}, Chen et~al. \cite{chen80}, Casteleiro et~al. \cite{casteleiro}.}
\end{figure*}

\begin{figure*}[!htb]
\centering
\includegraphics[width=0.85\textwidth]{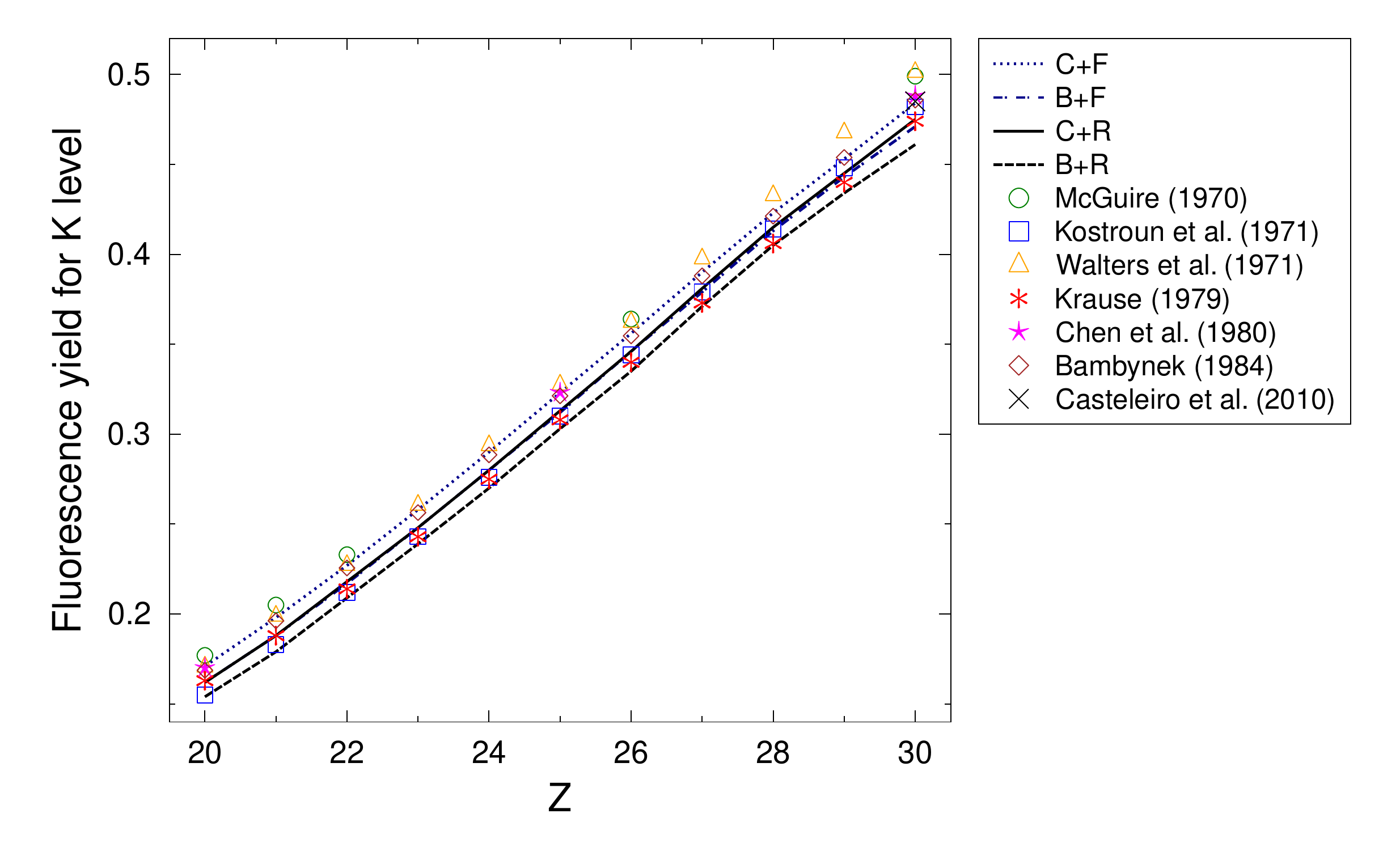}
\caption{\label{fig:komegapor}(Color online) Fluorescence yields of $K^{-1}$ hole state atomic level for atoms with 20$\le$Z$\le$30 compared to available theoretical and semi-empirical literature data: McGuire \cite{mcguire70}, Kostroun et~al. \cite{kostroun}, Walters et~al. \cite{walters}, Krause \cite{krause-omegi}, Chen et~al. \cite{chen80}, Bambynek \cite{bambynek} (obtained from the compilation of Hubbell \cite{hubbell}), Casteleiro et~al. \cite{casteleiro}.}
\end{figure*}

Figures \ref{fig:kwidpor} and \ref{fig:komegapor} present the comparison of data available in literature (theoretical and semi-empirical) discussed by other authors and present calculated values of $K^{-1}$ level width and $K^{-1}$ fluorescence yields, respectively.

\begin{table*}[!htb]
\caption{\label{tab:komegaporex}Fluorescence yields of $K^{-1}$ hole state atomic level for atoms with 20$\le$Z$\le$30 compared to last obtained experimental literature data.}
\medskip
\begin{tabular*}{\linewidth}{@{\extracolsep{\fill}}c cccc cccc}
\hline\\[-2.0ex]
Z & \multicolumn{4}{c}{Calculation model} & \multicolumn{4}{c}{Experimental data}\\[0.5ex]
\cline{2-5}\cline{6-9}\\[-2ex]
& C+R & B+R & C+F & B+F & Yashoda et al. & Durak and {\"O}zdemır & Gudennavar et al. & {\c S}im{\c s}ek et al. \\
&&&&& \cite{yashoda} & \cite{durak} & \cite{gudennavar} & \cite{simsek}\\[0.5ex]\hline\\[-1.5ex]
20 & 0.154 & 0.161 & 0.163 & 0.170 & -   & -   & -   & -   \\
21 & 0.184 & 0.175 & 0.193 & 0.184 & -   & -   & -   & -   \\
22 & 0.218 & 0.209 & 0.227 & 0.217 & 0.218 $\pm$ 0.008 & -   & -   & 0.214 $\pm$ 0.004 \\
23 & 0.248 & 0.239 & 0.258 & 0.248 & 0.249 $\pm$ 0.009 & -   & -   & 0.240 $\pm$ 0.008 \\
24 & 0.280 & 0.270 & 0.290 & 0.280 & -   & -   & -   & 0.291 $\pm$ 0.006 \\
25 & 0.313 & 0.303 & 0.323 & 0.312 & -   & 0.354 $\pm$ 0.007 & -   & 0.311 $\pm$ 0.008 \\
26 & 0.345 & 0.335 & 0.356 & 0.345 & -   & 0.330 $\pm$ 0.005 & -   & 0.331 $\pm$ 0.012 \\
27 & 0.381 & 0.371 & 0.389 & 0.379 & 0.375 $\pm$ 0.014 & -   & -   & 0.355 $\pm$ 0.011 \\
28 & 0.414 & 0.404 & 0.422 & 0.412 & 0.408 $\pm$ 0.015 & 0.412 $\pm$ 0.015 & -   & 0.448 $\pm$ 0.014 \\
29 & 0.440 & 0.448 & 0.448 & 0.456 & 0.438 $\pm$ 0.016 & 0.412 $\pm$ 0.029 & -   & 0.455 $\pm$ 0.015 \\
30 & 0.470 & 0.478 & 0.480 & 0.487 & 0.471 $\pm$ 0.018 & 0.482 $\pm$ 0.032 & 0.464 $\pm$ 0.010 & 0.482 $\pm$ 0.022 \\[0.5ex]\hline\\[-2.0ex]
Z & \multicolumn{4}{c}{Calculation model} & \multicolumn{3}{c}{Experimental data}&\\[0.5ex]
\cline{2-5}\cline{6-8}\\[-2ex]
& C+R & B+R & C+F & B+F & {\c S}ahin et al. & S{\"o}{\u g}{\"u}t & Han et al. & \\
&&&&& \cite{sahin} & \cite{sogut} & \cite{han} &\\[0.5ex]\hline\\[-1.5ex]
20 & 0.154 & 0.161 & 0.163 & 0.170 & 0.174 $\pm$ 0.021 & -   & -   &    \\
21 & 0.184 & 0.175 & 0.193 & 0.184 & -   & -   & -   &    \\
22 & 0.218 & 0.209 & 0.227 & 0.217 & 0.222 $\pm$ 0.027 & -   & 0.234 $\pm$ 0.019 &    \\
23 & 0.248 & 0.239 & 0.258 & 0.248 & 0.261 $\pm$ 0.031 & -   & 0.264 $\pm$ 0.021 &    \\
24 & 0.280 & 0.270 & 0.290 & 0.280 & -   & 0.265 $\pm$ 0.026 & 0.295 $\pm$ 0.024 &    \\
25 & 0.313 & 0.303 & 0.323 & 0.312 & -   & 0.297 $\pm$ 0.030 & 0.351 $\pm$ 0.028 &    \\
26 & 0.345 & 0.335 & 0.356 & 0.345 & -   & 0.366 $\pm$ 0.033 & 0.358 $\pm$ 0.029 &    \\
27 & 0.381 & 0.371 & 0.389 & 0.379 & -   & 0.391 $\pm$ 0.039 & -   &    \\
28 & 0.414 & 0.404 & 0.422 & 0.412 & -   & 0.451 $\pm$ 0.045 & 0.435 $\pm$ 0.035 &    \\
29 & 0.440 & 0.448 & 0.448 & 0.456 & -   & 0.478 $\pm$ 0.047 & 0.452 $\pm$ 0.036 &    \\
30 & 0.470 & 0.478 & 0.480 & 0.487 & -   & 0.525 $\pm$ 0.050 & 0.477 $\pm$ 0.038 &    \\[0.5ex]
\hline
\end{tabular*}
\end{table*}

\begin{figure*}[!htb]
\centering
\includegraphics[width=0.85\textwidth]{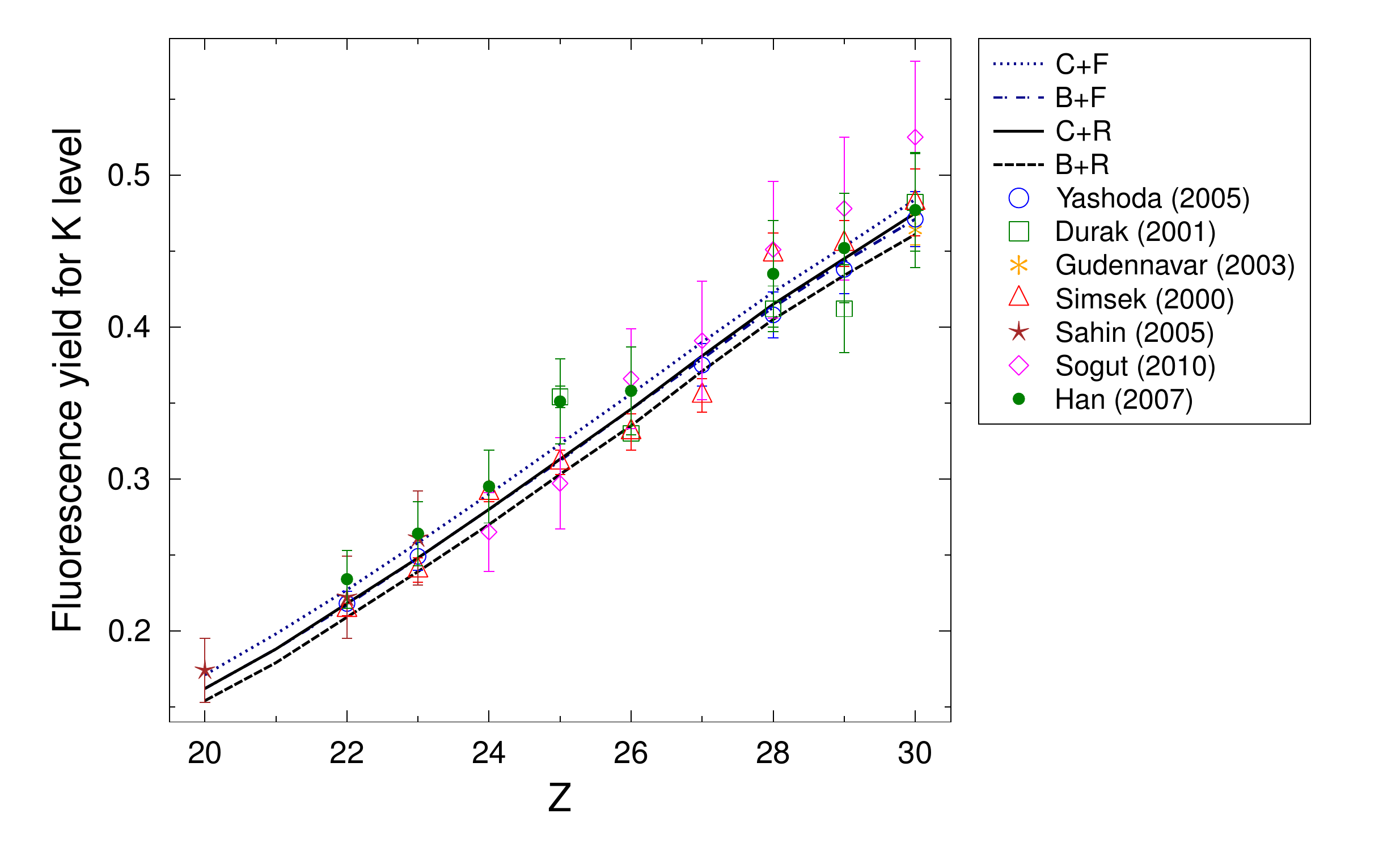}
\caption{\label{fig:komegaporex}(Color online) Fluorescence yields of $K^{-1}$ hole state atomic level for atoms with 20$\le$Z$\le$30 compared to last obtained experimental literature data: Yashoda et al. \cite{yashoda}, Durak and {\"O}zdemır \cite{durak}, Gudennavar et al. \cite{gudennavar}, {\c S}im{\c s}ek et al. \cite{simsek}, {\c S}ahin et al. \cite{sahin}, S{\"o}{\u g}{\"u}t \cite{sogut}, Han et al. \cite{han}.}
\end{figure*}

In Table~\ref{tab:komegaporex} and Figure~\ref{fig:komegaporex} the comparison between present calculated values of $K^{-1}$ fluorescence yields and the last years literature experimental data is presented. 
A good agreement between present calculated values and the other literature data is obtained, nevertheless, because of the significant differences in experimental data, it is hard to state which calculation model out of those mentioned above is closer fitted to experimental values of fluorescence yields. 
In order to obtain this statement, the residual variances (i.e. a sum of squared residuals divided by a number of considered experimental points) for experimental points and theoretical points of four models have been calculated. 
These variances are collected in Table~\ref{tab:lsqr}. 
From Tables~\ref{tab:komegaporex} and~\ref{tab:lsqr}, one can see that the experimental fluorescence yields presented in the papers of Yashoda et al. \cite{yashoda}, Durak and {\"O}zdemır \cite{durak}, and Gudennavar et al. \cite{gudennavar} are closer the theoretical values obtained by C+R model, the ones presented in the papers of {\c S}im{\c s}ek et al. \cite{simsek} and S{\"o}{\u g}{\"u}t \cite{sogut} are closer to B+F model, and finally, these presented in the papers of {\c S}ahin et al. \cite{sahin} and Han et al. \cite{han} are closer to C+F model. 
However, the experimental fluorescence yields obtained with the smallest error, i.e. Yashoda et al. \cite{yashoda} and Gudennavar et al. \cite{gudennavar} data, are very close to the theoretical values obtained by C+R model. 
This would suggest that relativistic calculations based on both MCDF-originated radiative and non-radiative transition rates reproduce experimental fluorescence yields values better.

\begin{table*}[!htb]
\caption{\label{tab:lsqr}Residual variances for experimental points and theoretical points of four models. All numbers are multiplied by $10^4$, e.g. 1.225 means 1.225$\times10^{-4}$.}
\medskip
\begin{tabular*}{\linewidth}{@{\extracolsep{\fill}}c ccccccc}
\hline\\[-1.5ex]
& Yashoda et al. & Durak and {\"O}zdemır & Gudennavar et al. & {\c S}im{\c s}ek et al. & {\c S}ahin et al. & S{\"o}{\u g}{\"u}t & Han et al. \\
Model & \cite{yashoda} & \cite{durak} & \cite{gudennavar} & \cite{simsek} & \cite{sahin} & \cite{sogut} & \cite{han} \\[0.5ex]\hline\\[-1.5ex]
C+R & 0.130 & 5.676 & 0.360 & 2.891 & 1.970 & 9.800 & 3.730 \\
B+R & 0.603 & 8.004 & 1.960 & 3.116 & 2.706 & 9.629 & 7.107 \\
C+F & 1.225 & 6.074 & 2.560 & 3.498 & 0.551 & 7.387 & 1.365 \\
B+F & 1.023 & 7.900 & 5.290 & 2.543 & 0.676 & 6.406 & 3.881 \\[0.5ex]
\hline
\end{tabular*}
\end{table*}

\section{Conclusions}
The $K$-shell level widths and $K$-shell fluorescence yields for atoms with 20$\le$Z$\le$30 have been calculated in relativistic way using MCDF and DHS calculations. 
The comparison between experimental fluorescence yields and theoretical ones, calculated by means of four models, results in satisfactory agreement. 
The present obtained results can be helpful in interpretation of experimental data of fluorescence yields and $K$-shell X-ray lines widths.

\subsection*{Acknowledgments}
The Author is thankful to Gustavo Aucar (Northeastern University of Argentina) for hosting during this paper was being prepared.

%

\begin{thebibliography}{38}
\expandafter\ifx\csname natexlab\endcsname\relax\def\natexlab#1{#1}\fi
\expandafter\ifx\csname bibnamefont\endcsname\relax
  \def\bibnamefont#1{#1}\fi
\expandafter\ifx\csname bibfnamefont\endcsname\relax
  \def\bibfnamefont#1{#1}\fi
\expandafter\ifx\csname citenamefont\endcsname\relax
  \def\citenamefont#1{#1}\fi
\expandafter\ifx\csname url\endcsname\relax
  \def\url#1{\texttt{#1}}\fi
\expandafter\ifx\csname urlprefix\endcsname\relax\def\urlprefix{URL }\fi
\providecommand{\bibinfo}[2]{#2}
\providecommand{\eprint}[2][]{\url{#2}}

\bibitem[{\citenamefont{McGuire}(1969)}]{mcguire69}
\bibinfo{author}{\bibfnamefont{E.~J.} \bibnamefont{McGuire}},
  \bibinfo{journal}{Phys. Rev. A} \textbf{\bibinfo{volume}{85}},
  \bibinfo{pages}{1} (\bibinfo{year}{1969}).

\bibitem[{\citenamefont{McGuire}(1970)}]{mcguire70}
\bibinfo{author}{\bibfnamefont{E.~J.} \bibnamefont{McGuire}},
  \bibinfo{journal}{Phys. Rev. A} \textbf{\bibinfo{volume}{2}},
  \bibinfo{pages}{273} (\bibinfo{year}{1970}).

\bibitem[{\citenamefont{Walters and Bhalla}(1971)}]{walters}
\bibinfo{author}{\bibfnamefont{D.~L.} \bibnamefont{Walters}} \bibnamefont{and}
  \bibinfo{author}{\bibfnamefont{C.~P.} \bibnamefont{Bhalla}},
  \bibinfo{journal}{Phys. Rev. A} \textbf{\bibinfo{volume}{3}},
  \bibinfo{pages}{1919} (\bibinfo{year}{1971}).

\bibitem[{\citenamefont{Kostroun et~al.}(1971)\citenamefont{Kostroun, Chen, and
  Craseman}}]{kostroun}
\bibinfo{author}{\bibfnamefont{V.~O.} \bibnamefont{Kostroun}},
  \bibinfo{author}{\bibfnamefont{M.~H.} \bibnamefont{Chen}}, \bibnamefont{and}
  \bibinfo{author}{\bibfnamefont{B.}~\bibnamefont{Craseman}},
  \bibinfo{journal}{Phys. Rev. A} \textbf{\bibinfo{volume}{3}},
  \bibinfo{pages}{533} (\bibinfo{year}{1971}).

\bibitem[{\citenamefont{Bambynek et~al.}(1972)\citenamefont{Bambynek,
  Crasemann, Fink, Freund, Mark, Swift, Price, and Rao}}]{bambynek72}
\bibinfo{author}{\bibfnamefont{W.}~\bibnamefont{Bambynek}},
  \bibinfo{author}{\bibfnamefont{B.}~\bibnamefont{Crasemann}},
  \bibinfo{author}{\bibfnamefont{R.W.} \bibnamefont{Fink}},
  \bibinfo{author}{\bibfnamefont{H.U.} \bibnamefont{Freund}},
  \bibinfo{author}{\bibfnamefont{H.}~\bibnamefont{Mark}},
  \bibinfo{author}{\bibfnamefont{C.D.} \bibnamefont{Swift}},
  \bibinfo{author}{\bibfnamefont{R.E.} \bibnamefont{Price}}, \bibnamefont{and}
  \bibinfo{author}{\bibfnamefont{P.V.} \bibnamefont{Rao}},
  \bibinfo{journal}{Rev. Mod. Phys.} \textbf{\bibinfo{volume}{44}},
  \bibinfo{pages}{716} (\bibinfo{year}{1972}).

\bibitem[{\citenamefont{Krause and Oliver}(1979)}]{krause}
\bibinfo{author}{\bibfnamefont{M.~O.} \bibnamefont{Krause}} \bibnamefont{and}
  \bibinfo{author}{\bibfnamefont{J.~H.} \bibnamefont{Oliver}},
  \bibinfo{journal}{J. Phys. Chem. Ref. Data} \textbf{\bibinfo{volume}{8}},
  \bibinfo{pages}{329} (\bibinfo{year}{1979}).

\bibitem[{\citenamefont{Krause}(1979)}]{krause-omegi}
\bibinfo{author}{\bibfnamefont{M.~O.} \bibnamefont{Krause}},
  \bibinfo{journal}{J. Phys. Chem. Ref. Data} \textbf{\bibinfo{volume}{8}},
  \bibinfo{pages}{307} (\bibinfo{year}{1979}).

\bibitem[{\citenamefont{Chen et~al.}(1980)\citenamefont{Chen, Crasemann, and
  Mark}}]{chen80}
\bibinfo{author}{\bibfnamefont{M.~H.} \bibnamefont{Chen}},
  \bibinfo{author}{\bibfnamefont{B.}~\bibnamefont{Crasemann}},
  \bibnamefont{and} \bibinfo{author}{\bibfnamefont{H.}~\bibnamefont{Mark}},
  \bibinfo{journal}{Phys. Rev. A} \textbf{\bibinfo{volume}{21}},
  \bibinfo{pages}{436} (\bibinfo{year}{1980}).

\bibitem[{\citenamefont{Hubbell et~al.}(1994)\citenamefont{Hubbell, Trehan,
  Singh, Chand, Mehta, Garg, Garg, Singh, and Puri}}]{hubbell}
\bibinfo{author}{\bibfnamefont{J.~H.} \bibnamefont{Hubbell}},
  \bibinfo{author}{\bibfnamefont{P.~N.} \bibnamefont{Trehan}},
  \bibinfo{author}{\bibfnamefont{N.}~\bibnamefont{Singh}},
  \bibinfo{author}{\bibfnamefont{B.}~\bibnamefont{Chand}},
  \bibinfo{author}{\bibfnamefont{D.}~\bibnamefont{Mehta}},
  \bibinfo{author}{\bibfnamefont{M.~L.} \bibnamefont{Garg}},
  \bibinfo{author}{\bibfnamefont{R.~R.} \bibnamefont{Garg}},
  \bibinfo{author}{\bibfnamefont{S.}~\bibnamefont{Singh}}, \bibnamefont{and}
  \bibinfo{author}{\bibfnamefont{S.}~\bibnamefont{Puri}}, \bibinfo{journal}{J.
  Phys. Chem. Ref. Data} \textbf{\bibinfo{volume}{23}}, \bibinfo{pages}{339}
  (\bibinfo{year}{1994}).

\bibitem[{\citenamefont{Campbell and Papp}(2001)}]{papp}
\bibinfo{author}{\bibfnamefont{J.~L.} \bibnamefont{Campbell}} \bibnamefont{and}
  \bibinfo{author}{\bibfnamefont{T.}~\bibnamefont{Papp}},
  \bibinfo{journal}{Atomic Data and Nuclear Data Tables}
  \textbf{\bibinfo{volume}{77}}, \bibinfo{pages}{1} (\bibinfo{year}{2001}).

\bibitem[{\citenamefont{Perkins et~al.}(1991)\citenamefont{Perkins, Cullen,
  Chen, Hubbell, Rathkopf, and Scofield}}]{eadl}
\bibinfo{author}{\bibfnamefont{S.~T.} \bibnamefont{Perkins}},
  \bibinfo{author}{\bibfnamefont{D.~E.} \bibnamefont{Cullen}},
  \bibinfo{author}{\bibfnamefont{M.-H.} \bibnamefont{Chen}},
  \bibinfo{author}{\bibfnamefont{J.~H.} \bibnamefont{Hubbell}},
  \bibinfo{author}{\bibfnamefont{J.}~\bibnamefont{Rathkopf}}, \bibnamefont{and}
  \bibinfo{author}{\bibfnamefont{J.~H.} \bibnamefont{Scofield}},
  \bibinfo{type}{Tech. Rep.} \bibinfo{number}{UCRL-50400 30},
  \bibinfo{institution}{Lawrence Livermore National Laboratory}
  (\bibinfo{year}{1991}).

\bibitem[{\citenamefont{Polasik et~al.}(2011)\citenamefont{Polasik,
  S{\l}abkowska, Rzadkiewicz, Kozio{\l}, Starosta, Wiatrowska-Kozio{\l},
  Dousse, and Hoszowska}}]{lifetimes-prl}
\bibinfo{author}{\bibfnamefont{M.}~\bibnamefont{Polasik}},
  \bibinfo{author}{\bibfnamefont{K.}~\bibnamefont{S{\l}abkowska}},
  \bibinfo{author}{\bibfnamefont{J.}~\bibnamefont{Rzadkiewicz}},
  \bibinfo{author}{\bibfnamefont{K.}~\bibnamefont{Kozio{\l}}},
  \bibinfo{author}{\bibfnamefont{J.}~\bibnamefont{Starosta}},
  \bibinfo{author}{\bibfnamefont{E.}~\bibnamefont{Wiatrowska-Kozio{\l}}},
  \bibinfo{author}{\bibfnamefont{J.-Cl.} \bibnamefont{Dousse}},
  \bibnamefont{and}
  \bibinfo{author}{\bibfnamefont{J.}~\bibnamefont{Hoszowska}},
  \bibinfo{journal}{Phys. Rev. Lett.} \textbf{\bibinfo{volume}{107}},
  \bibinfo{pages}{073001} (\bibinfo{year}{2011}).

\bibitem[{\citenamefont{Yashoda et~al.}(2005)\citenamefont{Yashoda,
  Krishnaveni, and Gowda}}]{yashoda}
\bibinfo{author}{\bibfnamefont{T.}~\bibnamefont{Yashoda}},
  \bibinfo{author}{\bibfnamefont{S.}~\bibnamefont{Krishnaveni}},
  \bibnamefont{and} \bibinfo{author}{\bibfnamefont{R.}~\bibnamefont{Gowda}},
  \bibinfo{journal}{Nucl. Instru. Meth. Phys. B}
  \textbf{\bibinfo{volume}{240}}, \bibinfo{pages}{607 } (\bibinfo{year}{2005}).

\bibitem[{\citenamefont{{\c S}ahin et~al.}(2005)\citenamefont{{\c S}ahin,
  Demir, and Budak}}]{sahin}
\bibinfo{author}{\bibfnamefont{M.}~\bibnamefont{{\c S}ahin}},
  \bibinfo{author}{\bibfnamefont{L.}~\bibnamefont{Demir}}, \bibnamefont{and}
  \bibinfo{author}{\bibfnamefont{G.}~\bibnamefont{Budak}},
  \bibinfo{journal}{Appl. Radiat. Isot.} \textbf{\bibinfo{volume}{63}},
  \bibinfo{pages}{141 } (\bibinfo{year}{2005}).

\bibitem[{\citenamefont{Gudennavar et~al.}(2003)\citenamefont{Gudennavar,
  Badiger, Thontadarya, and Hanumaiah}}]{gudennavar}
\bibinfo{author}{\bibfnamefont{S.B.} \bibnamefont{Gudennavar}},
  \bibinfo{author}{\bibfnamefont{N.M.} \bibnamefont{Badiger}},
  \bibinfo{author}{\bibfnamefont{S.R.} \bibnamefont{Thontadarya}},
  \bibnamefont{and}
  \bibinfo{author}{\bibfnamefont{B.}~\bibnamefont{Hanumaiah}},
  \bibinfo{journal}{Rad. Phys. and Chem.} \textbf{\bibinfo{volume}{68}},
  \bibinfo{pages}{721 } (\bibinfo{year}{2003}).

\bibitem[{\citenamefont{Durak and {\"O}zdemır}(2001)}]{durak}
\bibinfo{author}{\bibfnamefont{R.}~\bibnamefont{Durak}} \bibnamefont{and}
  \bibinfo{author}{\bibfnamefont{Y.}~\bibnamefont{{\"O}zdemır}},
  \bibinfo{journal}{Rad. Phys. Chem.} \textbf{\bibinfo{volume}{61}},
  \bibinfo{pages}{19 } (\bibinfo{year}{2001}).

\bibitem[{\citenamefont{{\c S}im{\c s}ek et~al.}(2000)\citenamefont{{\c S}im{\c
  s}ek, Do{\u g}an, Turgut, and Ertu{\u g}rul}}]{simsek}
\bibinfo{author}{\bibfnamefont{{\"O}.}~\bibnamefont{{\c S}im{\c s}ek}},
  \bibinfo{author}{\bibfnamefont{O.}~\bibnamefont{Do{\u g}an}},
  \bibinfo{author}{\bibfnamefont{{\"U}.}~\bibnamefont{Turgut}},
  \bibnamefont{and} \bibinfo{author}{\bibfnamefont{M.}~\bibnamefont{Ertu{\u
  g}rul}}, \bibinfo{journal}{Rad. Phys. Chem.} \textbf{\bibinfo{volume}{58}},
  \bibinfo{pages}{207 } (\bibinfo{year}{2000}).

\bibitem[{\citenamefont{S{\"o}{\u g}{\"u}t}(2010)}]{sogut}
\bibinfo{author}{\bibfnamefont{{\"O}.}~\bibnamefont{S{\"o}{\u g}{\"u}t}},
  \bibinfo{journal}{Chin. J. Phys.} \textbf{\bibinfo{volume}{48}},
  \bibinfo{pages}{212 } (\bibinfo{year}{2010}).

\bibitem[{\citenamefont{Han et~al.}(2007)\citenamefont{Han, {\c S}ahin, Demir,
  and {\c S}ahin}}]{han}
\bibinfo{author}{\bibfnamefont{I.}~\bibnamefont{Han}},
  \bibinfo{author}{\bibfnamefont{M.}~\bibnamefont{{\c S}ahin}},
  \bibinfo{author}{\bibfnamefont{L.}~\bibnamefont{Demir}}, \bibnamefont{and}
  \bibinfo{author}{\bibfnamefont{Y.}~\bibnamefont{{\c S}ahin}},
  \bibinfo{journal}{Appl. Radiat. Isot.} \textbf{\bibinfo{volume}{65}},
  \bibinfo{pages}{669 } (\bibinfo{year}{2007}).

\bibitem[{\citenamefont{Grant et~al.}(1980{\natexlab{a}})\citenamefont{Grant,
  McKenzie, Norrington, Mayers, and Pyper}}]{gr1}
\bibinfo{author}{\bibfnamefont{I.~P.} \bibnamefont{Grant}},
  \bibinfo{author}{\bibfnamefont{B.~J.} \bibnamefont{McKenzie}},
  \bibinfo{author}{\bibfnamefont{P.~H.} \bibnamefont{Norrington}},
  \bibinfo{author}{\bibfnamefont{D.~F.} \bibnamefont{Mayers}},
  \bibnamefont{and} \bibinfo{author}{\bibfnamefont{N.C.} \bibnamefont{Pyper}},
  \bibinfo{journal}{Comput. Phys. Commun.} \textbf{\bibinfo{volume}{21}},
  \bibinfo{pages}{207} (\bibinfo{year}{1980}{\natexlab{a}}).

\bibitem[{\citenamefont{Grant et~al.}(1980{\natexlab{b}})\citenamefont{Grant,
  McKenzie, and Norrington}}]{gr2}
\bibinfo{author}{\bibfnamefont{I.~P.} \bibnamefont{Grant}},
  \bibinfo{author}{\bibfnamefont{B.~J.} \bibnamefont{McKenzie}},
  \bibnamefont{and} \bibinfo{author}{\bibfnamefont{P.~H.}
  \bibnamefont{Norrington}}, \bibinfo{journal}{Comput. Phys. Commun.}
  \textbf{\bibinfo{volume}{21}}, \bibinfo{pages}{233}
  (\bibinfo{year}{1980}{\natexlab{b}}).

\bibitem[{\citenamefont{Grant and J.McKenzie}(1980)}]{gr3}
\bibinfo{author}{\bibfnamefont{I.~P.} \bibnamefont{Grant}} \bibnamefont{and}
  \bibinfo{author}{\bibfnamefont{B.}~\bibnamefont{J.McKenzie}},
  \bibinfo{journal}{J. Phys. B} \textbf{\bibinfo{volume}{13}},
  \bibinfo{pages}{2671} (\bibinfo{year}{1980}).

\bibitem[{\citenamefont{Hata and Grant}(1983)}]{gr4}
\bibinfo{author}{\bibfnamefont{J.}~\bibnamefont{Hata}} \bibnamefont{and}
  \bibinfo{author}{\bibfnamefont{I.P.} \bibnamefont{Grant}},
  \bibinfo{journal}{J. Phys. B} \textbf{\bibinfo{volume}{16}},
  \bibinfo{pages}{3713} (\bibinfo{year}{1983}).

\bibitem[{\citenamefont{Grant}(1984)}]{gr5}
\bibinfo{author}{\bibfnamefont{I.P.} \bibnamefont{Grant}},
  \bibinfo{journal}{Int. J. Quantum Chem.} \textbf{\bibinfo{volume}{25}},
  \bibinfo{pages}{23} (\bibinfo{year}{1984}).

\bibitem[{\citenamefont{Dyall et~al.}(1989)\citenamefont{Dyall, Grant,
  Johonson, Parpia, and Plummer}}]{gr6}
\bibinfo{author}{\bibfnamefont{K.~G.} \bibnamefont{Dyall}},
  \bibinfo{author}{\bibfnamefont{I.~P.} \bibnamefont{Grant}},
  \bibinfo{author}{\bibfnamefont{C.~T.} \bibnamefont{Johonson}},
  \bibinfo{author}{\bibfnamefont{F.~A.} \bibnamefont{Parpia}},
  \bibnamefont{and} \bibinfo{author}{\bibfnamefont{E.~P.}
  \bibnamefont{Plummer}}, \bibinfo{journal}{Comput. Phys. Commun.}
  \textbf{\bibinfo{volume}{55}}, \bibinfo{pages}{425} (\bibinfo{year}{1989}).

\bibitem[{\citenamefont{Grant}(1974)}]{grant1}
\bibinfo{author}{\bibfnamefont{I.P.} \bibnamefont{Grant}}, \bibinfo{journal}{J.
  Phys. B} \textbf{\bibinfo{volume}{7}}, \bibinfo{pages}{1458}
  (\bibinfo{year}{1974}).

\bibitem[{\citenamefont{Grant}(1988)}]{grant2}
\bibinfo{author}{\bibfnamefont{I.P.} \bibnamefont{Grant}},
  \emph{\bibinfo{title}{Relativistic Effects in Atoms and Molecules, in:
  Methods in Computational Chemistry}}, vol.~\bibinfo{volume}{2}
  (\bibinfo{publisher}{Plenum}, \bibinfo{address}{New York},
  \bibinfo{year}{1988}).

\bibitem[{\citenamefont{Polasik}(1989)}]{polasik1989a}
\bibinfo{author}{\bibfnamefont{M.}~\bibnamefont{Polasik}},
  \bibinfo{journal}{Phys. Rev. A} \textbf{\bibinfo{volume}{39}},
  \bibinfo{pages}{616} (\bibinfo{year}{1989}).

\bibitem[{\citenamefont{Polasik}(1995)}]{polasik1995}
\bibinfo{author}{\bibfnamefont{M.}~\bibnamefont{Polasik}},
  \bibinfo{journal}{Phys. Rev. A} \textbf{\bibinfo{volume}{52}},
  \bibinfo{pages}{227} (\bibinfo{year}{1995}).

\bibitem[{\citenamefont{Gu}(2008)}]{fac2}
\bibinfo{author}{\bibfnamefont{M.~F.} \bibnamefont{Gu}}, \bibinfo{journal}{Can.
  J. Phys.} \textbf{\bibinfo{volume}{86}}, \bibinfo{pages}{675}
  (\bibinfo{year}{2008}).

\bibitem[{\citenamefont{J{\"o}nsson et~al.}(2007)\citenamefont{J{\"o}nsson, He,
  Fischer, and Grant}}]{grasp2k}
\bibinfo{author}{\bibfnamefont{P.}~\bibnamefont{J{\"o}nsson}},
  \bibinfo{author}{\bibfnamefont{X.}~\bibnamefont{He}},
  \bibinfo{author}{\bibfnamefont{C.~Froese} \bibnamefont{Fischer}},
  \bibnamefont{and} \bibinfo{author}{\bibfnamefont{I.P.} \bibnamefont{Grant}},
  \bibinfo{journal}{Comput. Phys. Commun.} \textbf{\bibinfo{volume}{177}},
  \bibinfo{pages}{597} (\bibinfo{year}{2007}).

\bibitem[{\citenamefont{Babushkin}(1964)}]{cech-bab}
\bibinfo{author}{\bibfnamefont{F.~A.} \bibnamefont{Babushkin}},
  \bibinfo{journal}{Acta Phys. Polon.} \textbf{\bibinfo{volume}{25}},
  \bibinfo{pages}{749} (\bibinfo{year}{1964}).

\bibitem[{\citenamefont{Gu}(2003)}]{fac1}
\bibinfo{author}{\bibfnamefont{M.~F.} \bibnamefont{Gu}},
  \bibinfo{journal}{Astrophys. J.} \textbf{\bibinfo{volume}{582}},
  \bibinfo{pages}{1241} (\bibinfo{year}{2003}).

\bibitem[{\citenamefont{Fritzsche}(2001)}]{ratip}
\bibinfo{author}{\bibfnamefont{S.}~\bibnamefont{Fritzsche}},
  \bibinfo{journal}{J. Elect. Spect. Rel. Phen.}
  \textbf{\bibinfo{volume}{114}}, \bibinfo{pages}{1155} (\bibinfo{year}{2001}).

\bibitem[{\citenamefont{Fritzsche}(2012)}]{ratip2012}
\bibinfo{author}{\bibfnamefont{S.}~\bibnamefont{Fritzsche}},
  \bibinfo{journal}{Comput. Phys. Commun.} \textbf{\bibinfo{volume}{183}},
  \bibinfo{pages}{1525} (\bibinfo{year}{2012}).

\bibitem[{\citenamefont{Parpia et~al.}(1996)\citenamefont{Parpia, Fischer, and
  Grant}}]{grasp92}
\bibinfo{author}{\bibfnamefont{F.~A.} \bibnamefont{Parpia}},
  \bibinfo{author}{\bibfnamefont{C.~Froese} \bibnamefont{Fischer}},
  \bibnamefont{and} \bibinfo{author}{\bibfnamefont{I.~P.} \bibnamefont{Grant}},
  \bibinfo{journal}{Comput. Phys. Commun.} \textbf{\bibinfo{volume}{94}},
  \bibinfo{pages}{249} (\bibinfo{year}{1996}).

\bibitem[{\citenamefont{Casteleiro et~al.}(2010)\citenamefont{Casteleiro,
  Parente, Indelicato, and Marques}}]{casteleiro}
\bibinfo{author}{\bibfnamefont{C.}~\bibnamefont{Casteleiro}},
  \bibinfo{author}{\bibfnamefont{F.}~\bibnamefont{Parente}},
  \bibinfo{author}{\bibfnamefont{P.}~\bibnamefont{Indelicato}},
  \bibnamefont{and} \bibinfo{author}{\bibfnamefont{J.~P.}
  \bibnamefont{Marques}}, \bibinfo{journal}{Eur. Phys. J. D}
  \textbf{\bibinfo{volume}{56}}, \bibinfo{pages}{1} (\bibinfo{year}{2010}).

\bibitem[{\citenamefont{Bambynek}(1984)}]{bambynek}
\bibinfo{author}{\bibfnamefont{W.}~\bibnamefont{Bambynek}},
  \emph{\bibinfo{title}{A New Evaluation of K-Shell Fluorescence Yields}}
  (\bibinfo{publisher}{X-84 Proc.}, \bibinfo{address}{Leipzig},
  \bibinfo{year}{1984}).

\end{thebibliography}

\end{document}